\documentclass[submission, Phys]{SciPost}

\hypersetup{
    colorlinks,
    linkcolor={red!50!black},
    citecolor={blue!50!black},
    urlcolor={blue!80!black}
}

\usepackage[bitstream-charter]{mathdesign}
\usepackage{xcolor}
\usepackage{amsmath}
\usepackage{physics}
\usepackage{float}

\DeclareSymbolFont{usualmathcal}{OMS}{cmsy}{m}{n}
\DeclareSymbolFontAlphabet{\mathcal}{usualmathcal}

\begin{document}

\begin{center}{\Large \textbf{
A rectangular loop interferometer for scalar optical computations and controlled generation of higher-order vector vortex modes using spin-orbit interaction of light
}}\end{center}

\begin{center}
Ram Nandan Kumar\textsuperscript{1$\star$},
Gaurav Verma\textsuperscript{1},
Subhasish Dutta Gupta \textsuperscript{1,2,3$\star$},
Nirmalya Ghosh \textsuperscript{1$\star$} and
Ayan Banerjee \textsuperscript{1$\star$}

\end{center}

\begin{center}
{\bf 1} Department of Physical Sciences, Indian Institute of Science Education and Research Kolkata, Mohanpur-741246, West Bengal, India
\\

{\bf 2}Department of Physics, Indian Institute of Technology, Jodhpur 342030, India
\\
{\bf 3}Tata Institute of Fundamental Research Hyderabad, India \\

${}^\star$ {\small \sf rnk17ip025@iiserkol.ac.in}
${}^\star$ {\small \sf sdg@tifrh.res.in}
${}^\star$ {\small \sf nghosh@iiserkol.ac.in}
${}^\star$ {\small \sf ayan@iiserkol.ac.in}

\end{center}



\section*{Abstract}
{\bf
We have developed a rectangular loop interferometer (RLI) that confines light in a rectangular path and facilitates various interesting applications. Such a device can yield the sum of numerous geometric series converging to different values between zero and one by the use of simple intra-cavity beam splitters - both polarization-independent and dependent. Losses -- principally due to alignment issues of the beam in the RLI -- limit the average accuracy of the series sum value to be between 90 - 98\% (taking into account losses at the optics), with the computation speed determined by the bandwidth and response time of the detectors. In addition, with a circularly polarized input Gaussian beam, and a combination of half-wave plate and q-plate inserted into the interferometer path, the device can generate a vortex beam that carries orbital angular momentum (OAM) of all orders of topological charge. The OAM is generated due to the spin-orbit interaction (SOI) of light, and the topological charge increases with each successive pass of the beam inside the interferometer. However, experimentally, only the third order of OAM could be measured since projecting out individual orders entailed a slight misalignment of the interferometer, which caused higher orders to go out of resonance. Furthermore, with input linear polarization, the device can generate a vector beam bearing a superposition of polarization states resembling the multipole expansion of a charge distribution. Even here, experimentally, we were able to quantify the polarization distribution up to the third order (quadrupole term) using a Stokes vector analysis of the vector beam, with the size of the polarization singularity region increasing as the polarization states evolve inside the interferometer. Our work demonstrates the ubiquitous nature of loop interferometers in modifying the scalar and vector properties of light to generate simple mathematical results and other complex but useful applications. 
}

\vspace{10pt}
\noindent\rule{\textwidth}{1pt}
\tableofcontents\thispagestyle{fancy}
\noindent\rule{\textwidth}{1pt}
\vspace{10pt}

\section{Introduction}
\label{sec:intro}

Interferometers serve as powerful tools utilized across a multitude of scientific disciplines, enabling the precise measurement of minute changes in physical properties -- ranging from the behavior of atoms and subatomic particles to the very structure of the universe itself. The Michelson interferometer, a cornerstone in the history of physics, was instrumental in determining the speed of light as constant in a vacuum, as well as discarding the concept of the luminiferous aether\cite{michelson1887relative}. Similarly, the Mach-Zehnder interferometer finds extensive use in optical communications \cite{shimizu2000single}, sensing applications \cite{li2012all, deng2011highly,gao2011plasmonic}, Bose--Einstein condensation \cite{berrada2013integrated}, quantum information processing \cite{wheeler2014quantum,nielsen2001quantum}, etc. Another famous configuration, namely the Hanbury Brown-Twiss (HBT) interferometer, was developed to measure the size of stars by quantifying the correlation of intensity fluctuations \cite{dorrah2022tunable,brown1956correlation,hanbury1979test}. Subsequently, it has also facilitated studies such as two-photon interferences in quantum measurements- \cite{rarity1990two,fano1961quantum}, comparisons between Bosons and Fermions \cite{jeltes2007comparison}, and even the quantum interference of a single particle with itself \cite{shimizu2000single,neder2007interference}. Overall, travelling wave interferometers have played a pivotal role in advancing the precision measurement of atomic energy levels \cite{banerjee2001high,banerjee2003precise}. However, Fabry-Perot interferometers, also known as standing wave interferometers, are of crucial importance as well. Besides comprising the core of lasers \cite{siegman1986lasers}, these are also employed in optical filtering\cite{hays2010hybrid}, frequency stabilization \cite{neumann2008tunable}, and the detection of gravitational waves \cite{barish1999ligo}.

Now, spherical mirror standing wave cavities support both longitudinal and transverse modes. Moreover, the transverse boundary conditions imposed upon light confined inside such cavities lead to the generation of a longitudinal component in the light field. The presence of such a longitudinal component in the electric field distribution gives rise to a transverse component in the Poynting vector, which results in very interesting manifestations of the intrinsic spin-orbit interaction (SOI) of light \cite{bliokh2015spin,saha2018transverse,bliokh2009geometrodynamics,bliokh2010angular}, and the spin-Hall effect \cite{onoda2004hall,kavokin2005optical,hosten2008observation,fu2019spin}, even though the magnitude of these effects is exceedingly low in the paraxial regime \cite{marrucci2011spin}. Enhancement of these effects can be achieved by tightly focusing light or scattering light by small particles \cite{bliokh2009geometrodynamics,zhao2007spin,kumar2022probing}, which can even cause other fascinating phenomena such as spin-dependent optical vortex generation 
\cite{brasselet2009optical,bliokh2015spin,schwartz2006conservation},  photonic spin-momentum locking \cite{bliokh2015spin,pal2020direct}, spin-direction locking \cite{petersen2014chiral,o2014spin,bliokh2015spin,lefier2015unidirectional,nayak2023spin}, and so on. However, the effect of an interferometer on the SOI of light remains to be studied. Interestingly, inserting inhomogeneous anisotropic media, such as patterned liquid crystals (LCs), into an interferometer may lead to strong extrinsic SOI effects (i.e., exact conversion of the spin angular momentum (SAM) into OAM) in a light beam within the paraxial region itself \cite{liberman1992spin,bliokh2015spin,marrucci2006optical,soni2013enhancing}, which may lead to diverse applications.  

On the same note, the SOI effects inside an interferometer would also be enhanced for structured vector and vortex beams, especially since these beams carry SAM, linked to the polarization properties, and OAM, associated with the helical phase front of light \cite{andrews2012angular,marrucci2006optical}. As a result, SOI creates intriguing effects within the framework of such vector and vortex beams due to their inhomogeneous
polarization distributions and phase singularity regions \cite{bliokh2015spin, shao2018spin, zhao2007spin,kumar2024probing}. These properties render  vector beams such as radially and azimuthally polarized beams useful in a variety of applications including multiplexing \cite{gregg2019enhanced}, the optical Hall effect \cite{li2015observation,Int_10, Int_11, Int_12, Int_13, kumar2022probing}, quantum sensing \cite{degen2017quantum, pirandola2018advances}, quantum information \cite{bennett2000quantum, ladd2010quantum}, optical communication \cite{ren2016orbital}, particle trapping \cite{padgett2011tweezers,kumar2022manipulation}, optical encryption \cite{liu2017single}, quantum memory \cite{nicolas2014quantum}, and super-resolution microscopy \cite{torok2004use}. It can thus be envisaged that augmenting SOI effects in such beams by circulating them in an interferometer can even lead to the construction of customized  vortex beams and vector beams that would possess space-varying 3D polarization properties \cite{bauer2015observation,quabis2005generation,cardano2012polarization,beresna2011radially,cardano2013generation}.

This is the focus of our study in this paper - where we have developed a rectangular loop interferometer (RLI) that can confine light in a rectangular path geometry using mirrors and beam splitters. First, we exploited the reflection and transmission properties of the optical elements in each successive cyclic path of the interferometer to generate the equivalent of a converging mathematical series, whose sum may be estimated by physical measurements at the output of the RLI. The estimation of the series sum is actually performed at the speed of light, though the rate of physical information collection is limited by the detector bandwidth and response time. Note that the operation of a standard Fabry-Perot interferometer also mimics the sum of a converging geometric series, but for such interferometers, the sum ideally always converges to 1 (neglecting losses at the mirrors). In our case, however - the RLI may be tuned to obtain solutions  of 1 or different from 1 by varying the intensity and/or the polarization of the circulating light, as we describe later.

Then, we proceeded to physically generate a OAM-carrying Laguerre-Gaussian or vortex beam containing the superposition of a mathematical series of such beams carrying OAM of all possible orders of topological charge except zero. Generally, a vortex half-wave retarder (q-plate) of zero-order with topological charge q can generate $2q\hbar$ OAM per photon \cite{marrucci2006optical}. However, in our case, we controlled the value of OAM simply by using a single q-plate ($q=1/2$) and a half-wave retarder or waveplate (HWP) into one arm of the RLI, and propagating a circularly polarized input Gaussian beam through these devices. As the circularly polarized Gaussian (RCP/LCP) beam passed through the q-plate, the resulting SOI of light\cite{marrucci2006optical, li2015observation} led to an increase in the value of OAM after every complete cycle, accompanied by an opposite helicity of SAM generated correspondingly.  Experimentally, we successfully measured the OAM value of topological charge ranging from $l$ = -3 to 3. To verify this, we superposed the output vortex beam from the RLI with a circularly polarized fundamental Gaussian so as to generate `fork' (or spiral) patterns, which demonstrated the topological charge carried by the beams. We deliberately misaligned the interferometer slightly to project out the individual OAM states, which - however, imposed a limitation on measuring higher order OAM states, which tended to leak out of the interferometer due to amplification of their misalignment within the RLI.

Finally, we generated a physical vector beam with an inhomogeneous complex polarization state that was again the superposition of a series of  polarization states of light. Indeed, the final polarization state may be considered as a multipole-like superposition (monopole, dipole, quadrupole, hexapole, octupole, etc.) of field polarization states. Thus, as the beams circulated within the interferometer RLI, the size of the polarization (intensity) singularity region increased, resulting in continuous changes in polarization within the beam. Theoretically, such multipole-like polarization states can be understood by employing the Jones vector algebra at different points of the RLI\cite{zhou2023polarization, davis2015analysis}. However, experimentally, we measured the monopole, dipole and quadrupole-like polarisation states by utilizing a Stokes vector analysis.  Here again, we could not measure higher order polarization states since we had to misalign the RLI again in order to project out the different states of polarization individually\cite{wolf2007introduction,gupta2015wave,azzam2016stokes}. 

\begin{center}
\begin{figure}
\includegraphics[width=\textwidth]{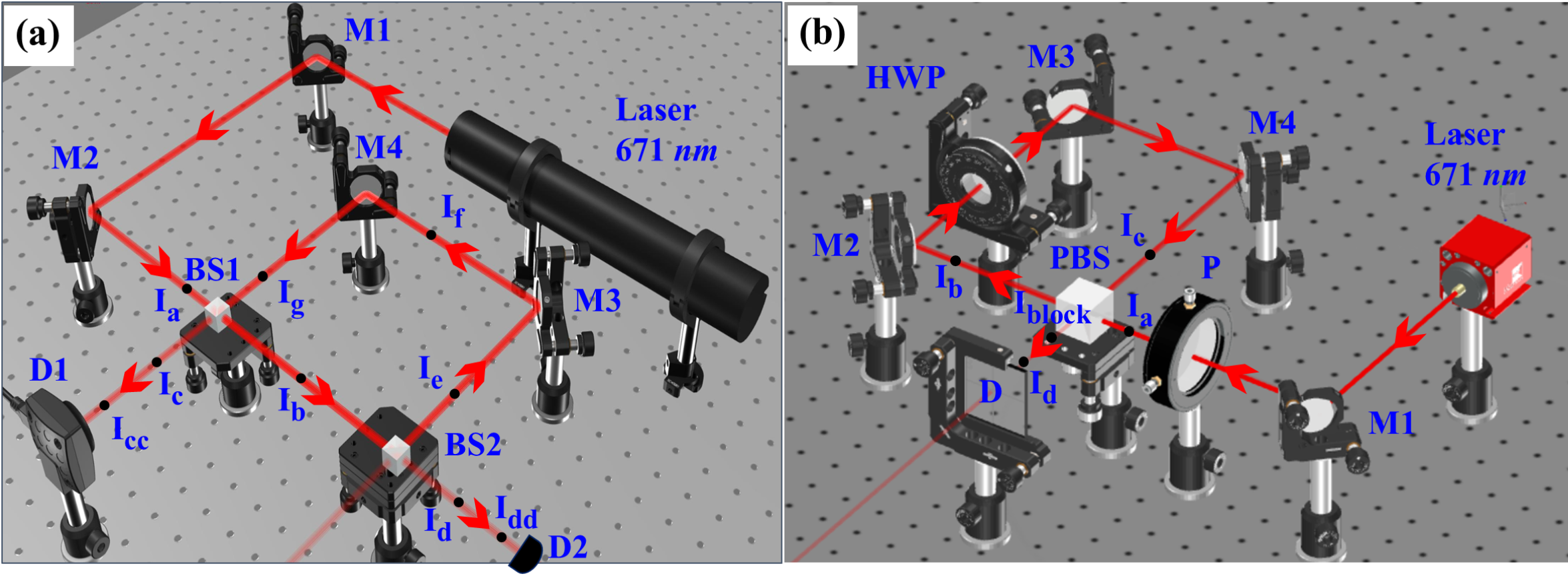}
\caption{Schematic diagrams of RLI setup, which provide the converging geometric series. (a) The intensity-dependent geometric series I$_{1}$ and I$_{2}$ are obtained in detectors D1 and D2, respectively. (b) The polarization-dependent geometric series I$_{p}$ is obtained at the output of the PBS (or detector D). These series converse to one for all values of $\theta$, where $\theta$ is the angle of rotation of the HWP fast-axis with respect to the polarization state of the input beam. BS1 and BS2 are the non-polarizing beam splitters; PBS is the polarizing beam splitter; M1, M2, M3, and M4 are the mirrors; P is the polarizer; D1, D2 and D are the detectors.}
\label{series sch}
\end{figure}
\end{center}

\section{Determining the sum of mathematical geometric series using the RLI adapting multiple approaches }

The RLI is a unidirectional path interferometer; on every complete circulation of the beam into the loop of the interferometer, the intensity reduces significantly depending on the number and nature of the beam splitters used in the RLI; therefore, the contrast of the fringes is not quite good (see Fig.~\ref{RLI} (b) of Appendix A). However, it has the advantage of adding a variety of optical elements in its path -- so that the interferometer is not limited to summing mathematical series which would converge to 1. Thus, we used the RLI in three different configurations to generate the sum of: (a)  geometric series using the variation of light intensity, (b) geometric series using the variation of light polarization, and (c) geometric series using the variation of both light intensity and polarization. We determined the accuracy of our measurements from the power we measured at the output of the interferometer, which we normalized with respect to the input power, and compared this value to the theoretical sum of the particular series. Thus, we obtained an average accuracy of around 90-98\% by taking into account the intensity losses from the mirrors and beam splitters. The residual error may have arisen from misalignment issues and limitations of the paraxial beam approximation. In what follows, we describe the theoretical and experimental premise of our work, with detailed descriptions of the three cases mentioned above.
\subsection{Case 1: Sum of a converging geometric series ($ \lesssim 1$) using the evolution of light intensity in the RLI}
As depicted in Fig.~\ref{series sch} (a), a fundamental Gaussian beam with linear polarization passed through a 20 mm cubic nonpolarizing beam splitter 50-50 (BS1). The beam underwent first-order transmission (T1) and reflection (R1) before the second nonpolarizing beam splitter 50-50 (BS2). The first-order reflected beam R1 was incident on the detector (D1). The transmitted beam T1 from BS1 again underwent first-time transmission (T2) and reflection (R2) from BS2. The T2 beam directly reached the detector (D2); however, the R2 beam was reflected from two mirrors of RLI in such a way that it was incident on the BS1. In BS1, the second-order transmission T3 fell on detector D1 and the reflection (R3) was incident on BS2. At BS2, the second-order transmitted beam T4 again reached the detector D2, while the reflection R4 went back into the loop. In this way, the beams repeated the same loop repeatedly. At detector D1,  we measured the intensity of the first order of reflection R1 and the odd orders of transmission (T3, T5, T7,…). However, at detector D2, we measured the intensity of even orders of transmission (T2, T4, T6,…) using the power meter. 

At detector D1, we may write the geometric series as
\begin{equation}
\begin{aligned}
I_1=I_a\left(\frac{1}{2^1}+\frac{1}{2^3}+\frac{1}{2^5}+\frac{1}{2^7}+\cdots\right)
\label{I1}
\end{aligned}
\end{equation}
where, $I_{a}$ is the input intensity of the beam (see Fig~\ref{series sch} (a)). The sum of the series would be $I_{1}=\frac{2}{3}I_{a}=1.98 $. However, experimentally we obtained  $I_{1}=1.703$ at detector D1. This is because our beam splitters, mirrors and detector are not ideal; therefore, as mentioned earlier, the intensity losses through optical components account for around 12\% of the error (see Table.~\ref{table1} of Appendix A). The residual error of 1.6\% was due to the limitations imposed by the paraxial beam approximation and alignment issues. 

For the geometric series formed at detector D2, we have
\begin{equation}
\begin{aligned}
I_2=I_a\left(\frac{1}{4^1}+\frac{1}{4^2}+\frac{1}{4^3}+\frac{1}{4^4}+\cdots\right)
\label{I2}
\end{aligned}
\end{equation}

The sum of the series would be $I_{2}=\frac{1}{3}I_{a}=0.9933$. However, we experimentally obtained $I_{2}=0.75$ at detector D2 (see Fig.~\ref{series sch} (a)), and we get that the error is around 24\%. However, when we took intensity losses due to BS's and mirrors into account, the error reduced to 6.4\%, and we obtained 94.6\% accuracy (see Table.~\ref{table2} of Appendix A).

We can obtain different geometric series by using different types of beam splitters at the vertices of RLI. However, the intensity losses through the optical components are the major sources of error in determining the sum of the geometric series accurately. Using a minimum number of optical components to confine the light into the loop appears better. Therefore, in Case `2', we developed a new method to generate geometric series using PBS and HWP (or quarter wave-plate (QWP)) to avoid such intensity losses.

\begin{figure}[!h]
    \begin{center}
    \includegraphics[width=0.5\textwidth]{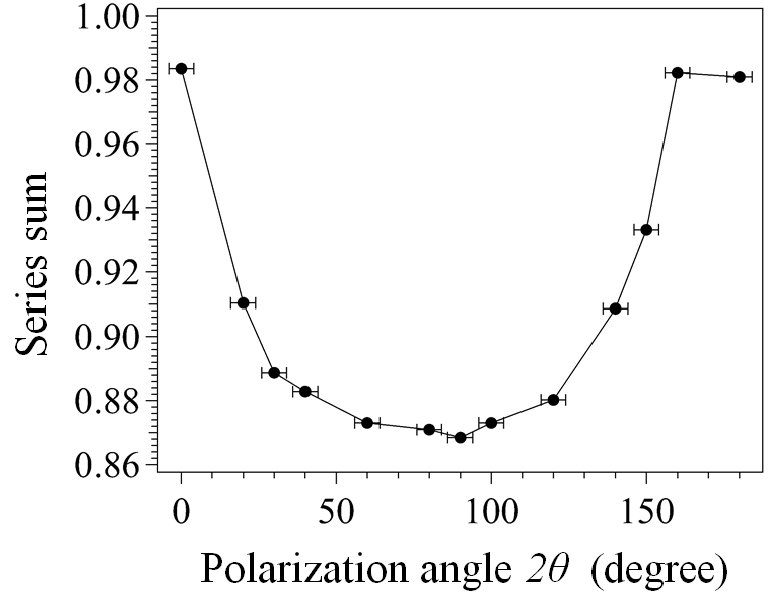}
    \caption{(a) The geometric series sum of the p-polarized state is a function of the polarization state generated by HWP into RLI.}
    \label{HWP}
    \end{center}
\end{figure}

\subsection{Case 2: Sum of a converging geometric series converging to $\sim 1$ using the evolution of polarization-dependent light intensity in the RLI} 
As illustrated in Fig.~\ref{series sch} (b), we now constructed the RLI using a combination of PBS, HWP and mirrors. The linearly p-polarized Gaussian beam was initially incident on the PBS (in Fig.~\ref{series sch} (b)). The transmitted T1 beam was p-polarized, and the reflected R1 was s-polarized. The intensity of the reflected beam R1 was not zero since our polarizer and PBS were not ideal (we considered this intensity as part of our error budget). The transmitted T1 beam (p-polarized) passed through the HWP. Understandably, the HWP rotated the input degree of polarization by an angle ($2\theta$) when a linearly p-polarized beam was incident at an angle $\theta$ with respect to the fast axis of the HWP. The beams emerging from HWP had both p and s components of polarization. After reflecting from three mirrors of RLI, the PBS only transmitted T2 (p-polarized beam) toward detector D and reflected R2 (s-polarized)  into the loop. Now, this time, the HWP again rotated the degree of polarization of the s-polarized beam by an angle of $2\theta$. The emerging wave from the HWP was again p- and s-polarized, and the same scenario repeated again and again. Our detector D measured the intensity of only the p-polarized beams transmitted from the PBS (P1+P2+P3+…). Most importantly, we could have different geometric series for different values of $\theta$, but the series converged to 1 for all values of $\theta$.

\begin{equation}
\begin{aligned}
P & =P_1+P_2+P_3+\cdots \\
I_P & =I_0\left(\cos ^2 2\theta+\sin ^4 2\theta \cdot \sum_{n=0}^{\infty} \cos ^{2 n} 2\theta\right); \forall 2\theta
\end{aligned}
\label{Ip}
\end{equation}\\

for $2\theta=30^{\circ}$
\begin{equation}
I_p=I_0\left(\frac{3}{4}+\frac{1}{16}\left(1+\frac{3}{4}+\frac{3^2}{4^2}+\frac{3^3}{4^3}+\cdots\right)\right)
\end{equation}

for $2\theta=60^{\circ}$
\begin{equation}
I_p=I_0\left(\frac{1}{4}+\frac{9}{16}\left(1+\frac{1}{2^2}+\frac{1}{2^4}+\frac{1}{2^6}+\cdots\right)\right)
\end{equation} \\

In Figure \ref{HWP}, we show the sum of the series represented by Eq.~ \ref{Ip} for different theta values of the HWP. As we increased the value of $\theta$ from $0^{\circ}$ to $45^{\circ}$, the s-polarized component increased in the interferometer path; therefore, the error also increased. At $\theta =45^{\circ}$, the series sum had a maximum error of around 13\% because the HWP converted the p-polarized component entirely into s-polarized. As mentioned earlier, the PBS reflected the s-polarized component of the beam into the loop; therefore, some of the s-polarized components remained even after an infinite number of circulations inside the RLI. At $0^{\circ}$ or $90^{\circ}$ of the fast axis of HWP, the beam was completely p-polarized; therefore, the beams could not traverse the loop after a single pass, so we obtained the minimum error of around 2\%.
\begin{figure}[!h]
    \begin{center}
    \includegraphics[width=\textwidth]{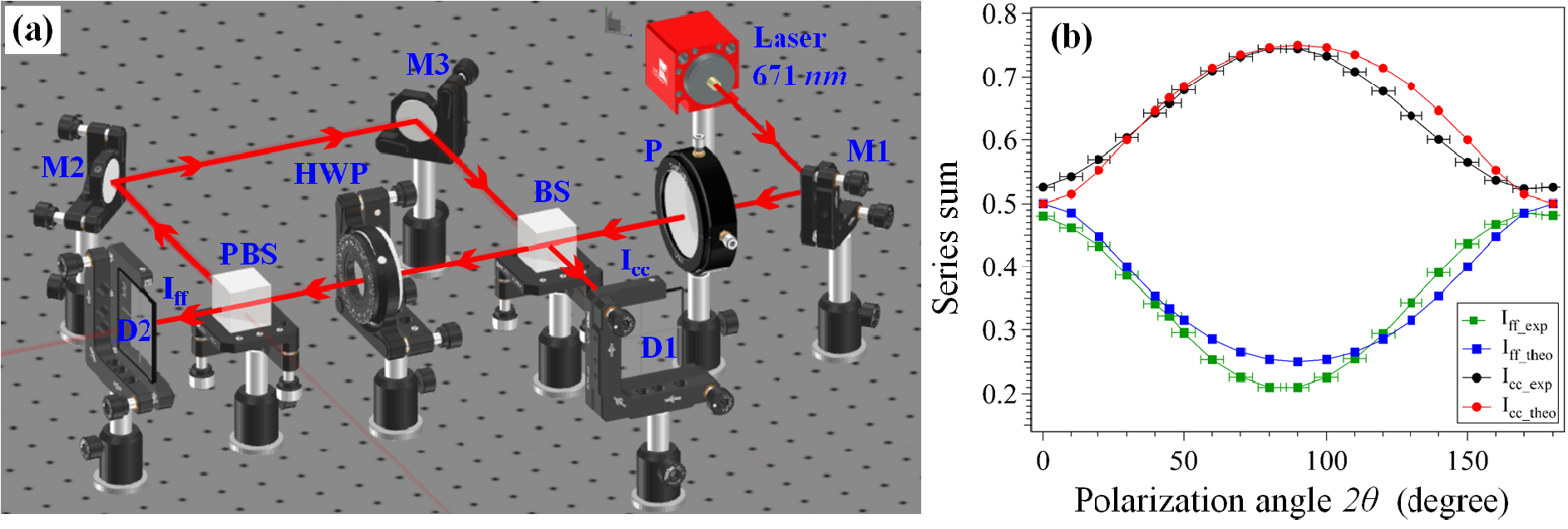}
    \caption{ (a) Schematic diagram of the RLI setup as mentioned in Case 3 that generates the converging geometric series (see Eq.~\ref{Variable series}). The geometric series obtained at the output of the first beam splitter (BS) $I_{cc}$ and second polarizing beam splitter (PBS) $I_{ff}$, respectively.  These series converge to all possible points lying between 0.25 to 0.75 depending upon the value of $2\theta$, where $\theta$ is the angle of rotation of HWP fast-axis with respect to the input polarization state of the Gaussian beam, M1, M2, M3 are the mirrors, P is the polarizer, D1 and D2 are the detectors. (b) The convergent geometric series sum is a function of the polarization state generated by HWP and PBS into RLI.}
    \label{BS_HWP_PBS}
    \end{center}
\end{figure}


\subsection{Case 3: Sum of a geometric series converging to various values using the evolution of polarization and intensity of light in the RLI} 
As shown in Fig.~\ref{BS_HWP_PBS} (a), we constructed the RLI with the help of the BS, PBS and two mirrors at the vertices of the rectangular geometry. Between BS and PBS, we inserted the HWP so that it rotated the polarization of the input beam. First, a p-polarized (horizontally polarized) Gaussian beam was incident on the non-polarizing BS (50-50). The reflected beam R1 was incident directly on the detector D1, and the transmitted T1 light passed through HWP. The fast-axis of the HWP was at $\theta$ degree with respect to the input polarization of the T1 beam, so the HWP rotated the degree of polarization by an angle $2\theta$. Therefore, the emerging beams from HWP had both p and s components of polarization. The p and s polarized components were transmitted and reflected from the PBS, respectively. The transmitted p-polarized component was incident on detector D2. However, the s-polarized component continued circulating in the loop of RLI. After every complete circulation of the beam into the loop, the s and p polarized components were collected at detectors D1 and D2, respectively. 

\begin{equation}
\begin{aligned}
& I_{c c}=I_0\left(\frac{1}{2}+\sin ^2 2 \theta \cdot \sum_{n=0}^{\infty} \frac{\cos ^{2 n} 2 \theta}{2^{n+2}}\right) \\
& I_{ff}=I_0\left(\frac{\cos ^2 2 \theta}{2}+\sin ^4 2 \theta \cdot \sum_{n=0}^{\infty} \frac{\cos ^{2 n} 2 \theta}{2^{n+2}}\right)
\end{aligned}
\label{Variable series}
\end{equation}

Here $I_{0}$ is the input intensity, $\theta$ is the angle of rotation of the HWP fast-axis with respect to the input polarization state of the Gaussian beam. Note that at $\theta =0^{\circ}$, the beam is p-polarized. $I_{cc}$ and $I_{ff}$ are the output intensity collected at the detector D1 and D2, respectively. These series converged to all possible points lying between 0.25 to 0.75 depending upon the value of $2\theta$. In Fig.~\ref{BS_HWP_PBS} (b), the solid red and black circles represent the theoretical and experimental values of the converging geometric series $I_{cc}$ lying between 0.5 and 0.75, respectively, at detector D1. However, the solid blue and green squares represent the theoretical and experimental values of the converging geometric series $I_{ff}$ lying between 0.25 and 0.50, respectively, at detector D2. At $\theta=45$, the emerging beam from the HWP was completely s-polarized; therefore, the series $I_{cc}$ and $I_{ff}$ attained maximum and minimum values of 0.75 and 0.25, respectively. Again, the minimum and maximum errors at detectors D1 ($I_{cc}$) and D2 ($I_{ff}$) at $\theta$ were around 40-45 degrees, respectively. We have also checked that if we alter the position of BS and PBS, the series formed at detectors D1 and D2 were also interchanged; that is, $I_{cc}$ become $I_{ff}$ and vice versa. Thus, it is possible to obtain geometric series converging to all possible values between 0 and 1 by varying the reflectance and transmittance of the BS.

\section{Generation of a complex vortex beam containing a superposition of topological charges using the RLI}
The most common solution of the paraxial wave equation is the fundamental Gaussian mode; however, a higher-order solution also exists depending on the symmetry of the coordinate system. The rectilinear and cylindrical symmetry provides the Hermite-Gaussian (HG) modes and Laguerre-Gaussian (LG) mode solutions, respectively \cite{andrews2012angular,zhan2009cylindrical}. The electric fields of the LG modes can be described in cylindrical coordinates $(r, \phi, z)$ as \cite{yao2011orbital,bierdz2013high} 

\begin{equation}
\begin{aligned}
L G_{p \ell}= & \sqrt{\frac{2 p !}{\pi(p+|\ell|) !}} \frac{1}{w(z)}\left[\frac{r \sqrt{2}}{w(z)}\right]^{|\ell|} \exp \left[\frac{-r^2}{w^2(z)}\right] L_p^{|\ell|}\left(\frac{2 r^2}{w^2(z)}\right) \exp [i \ell \phi] \\
& \exp \left[\frac{i k_0 r^2 z}{2\left(z^2+z_R^2\right)}\right] \exp \left[-i(2 p+|\ell|+1) \tan ^{-1}\left(\frac{z}{z_R}\right)\right]
\label{LG beam theo}
\end{aligned}
\end{equation}


where $w({z})$ is the beam waist at z distance from the focus, $w_{0}$ is the beam waist at the focus, $Z_{R}$ is the Rayleigh range, $L^{|l|}_{p} (x)$ is the Laguerre polynomial, $(2p+|l|+1)\tan[-1]({\frac{Z}{Z_{R}}})$ is the Gouy phase, $p > 0$ labels the radial modes, and $exp(i\ell\phi)$ is the helical phase of the beam $(\ell \hbar ~(\ell \in \mathbb{Z})$ is the OAM per photon). Some other beams carrying OAM with similar phase structures are the high-order Bessel \cite{durnin1987diffraction}, and Mathieu \cite{gutierrez2000alternative} beams. Until now, there are many ways of generating LG (OAM) beams, which include spiral phase plates \cite{oemrawsingh2005experimental}, spatial light modulators (SLM)\cite{yao2011orbital}, q-plates \cite{marrucci2006optical}, metamaterials \cite{dorrah2022tunable}, etc. Generally, a q-plate with a fixed charge value $q=1/2$ generates a first-order LG beam \cite{marrucci2006optical,nagali2009quantum}. However, here, we generate beams possessing higher-order OAM using a single q-plate ($q=1/2$) inside an interferometer. Further details are provided below.

\begin{center}
\begin{figure}
\includegraphics[width=\textwidth]{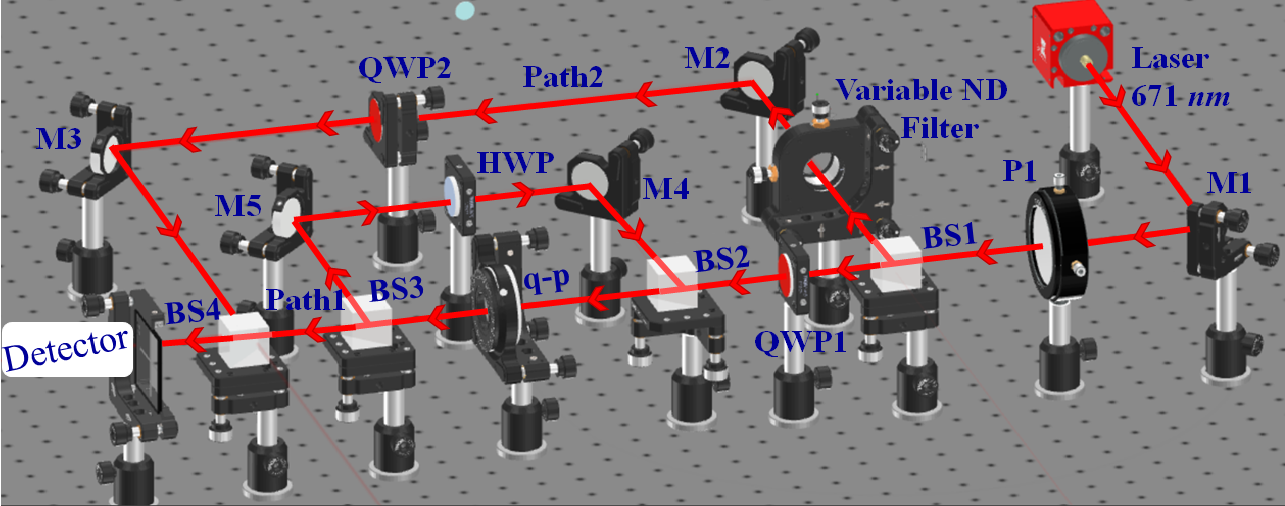}
\caption{Schematic diagram of our experimental setup used to produce a higher order $LG_{pl}$ (OAM) beam. Where M1, M2, M3, M4, and M5 are the mirrors, P1 is a wire grid polarizer, QWP1 and QWP2 are the quarter wave plate, HWP is a half-wave plate, q-p is q-plate, BS1, BS2, BS3 and BS4 are the 50-50 non-polarizing beam splitters.}
\label{OAM Schematic}
\end{figure}
\end{center}

A schematic of our experimental system is provided in Fig.~\ref{OAM Schematic}. The fundamental Gaussian mode of a solid-state laser (Lasever, 671 nm) was passed through a wire grid polarized P1, a 10 mm cubic beam splitter (BS1); therefore, we had two beam paths- transmission (path1) and reflection (path2) of the beam. The path1 beam passed through a quarter-wave plate (QWP$_{1}$) which was at $45^{o}/135^{o}$ with respect to the polarizer P$_{1}$. The circularly polarized (LCP/RCP) light in path 1 entered into an interferometer (RLI). The RLI was made of two beam splitters (BS2 and BS3, $50/50$) and two mirrors (M4 and M5). Inside the RLI, we used two more optical components, a q-plate ($q=1/2$) and a HWP, to generate LG beams. The q-plates are fabricated with fixed q-values having special liquid crystal patterns. These are variable half-wave retarders (HWR), where the principal axis spatially varies by an orientation angle given by $\psi(\phi) = q.\phi + \psi_{0}$. It is possible to add, subtract, or change the sign of the charge only by combining q-plates with HWPs. Generally, a q-plate with topological charge q can generate $2q\hbar$ angular momentum (AM) per photon. For the Jones matrix calculation of an inhomogeneous HWR, we assumed an incident left and right circularly polarized (LCP/RCP) plane wave, denoted as $\mathbf{E^{LCP/RCP}_{in} = [~1, ~~\pm i~~]^T}$. This assumption corresponds to the idealization associated with the central wave vector of a paraxial Gaussian beam, often referred to as $TM_{00}$. 

The output field emerging from the inhomogeneous HWR can be expressed as:
\begin{equation}
J_{H W R}(\delta, \psi)=R^{-1}(\psi) J_{H W R} R(\psi), 
\end{equation}
such that,
\begin{equation}
\begin{aligned}
& J_{H W R}(\delta, \psi)=\left[\begin{array}{cc}
\cos \psi & -\sin \psi \\
\sin \psi & \cos \psi
\end{array}\right]\left[\begin{array}{cc}
1 & 0 \\
0 & -1
\end{array}\right]\left[\begin{array}{cc}
\cos \psi & \sin \psi \\
-\sin \psi & \cos \psi
\end{array}\right]=\left[\begin{array}{cc}
\cos 2 \psi & \sin 2 \psi \\
\sin 2 \psi & -\cos 2 \psi
\end{array}\right] \\\\
& \mathbf{E^{RCP/LCP}_{out}}=J_{H W R}(\delta, \psi) {E^{LCP/RCP}_{in}}=\exp (2 i \psi)\left[\begin{array}{c}
1 \\
\mp i
\end{array}\right]=\exp (\pm 2 i q \phi) \times \exp \left(\pm 2 i \psi_0\right)\left[\begin{array}{c}
1 \\
\mp i
\end{array}\right]
\end{aligned}
\label{SOI}
\end{equation}



\begin{figure}[!h]
    \begin{center}
\includegraphics[width=\textwidth]{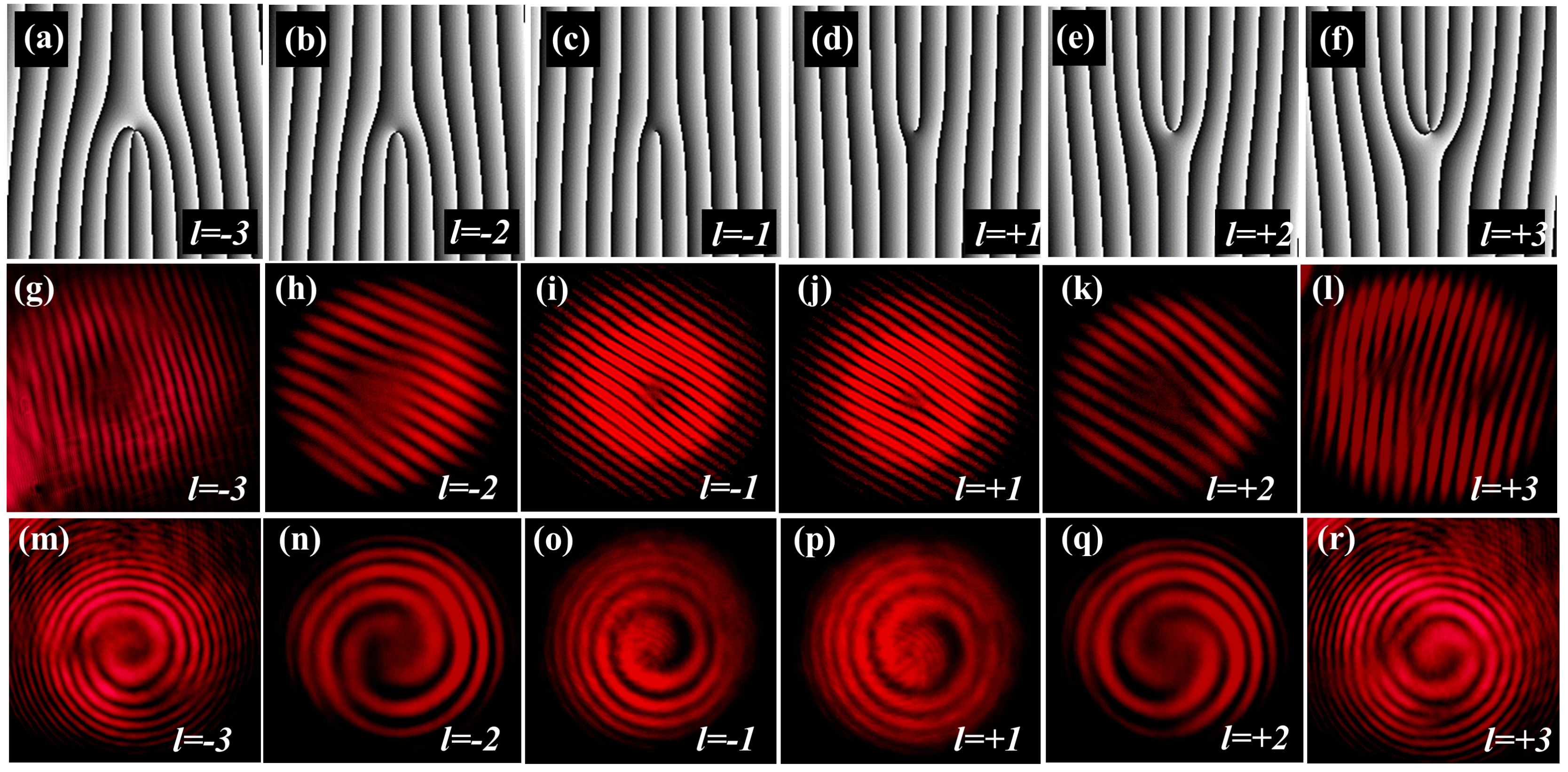}
\caption{The upper panels [(a)-(f)] depict the forked pattern representing the OAM state $l=-3$ to $l=+3$ of the LG beam. This pattern is created by combining the phase distribution of the LG ($l$) modes with a linear phase ramp, resulting in a forked pattern. The middle panels [(g)-(l)] show the interference patterns of the RCP/LCP helical (OAM) modes emerging from the q-plate ($q=1/2$) inside RLI when superimposed with a reference RCP/LCP Gaussian plane wavefront geometry (collimated). The lower panels [(m)-(r)] display the interference patterns of the RCP/LCP helical (OAM) modes when superimposed with the reference RCP/LCP Gaussian mode (which has a spherical wavefront after focusing by a lens). Figures on the left side [(g), (h), (i), (m), (n), and (o)] correspond to left-circular input polarization, while those on the right side [(j), (k), (l), (p), (q), and (r)] correspond to right-circular input polarization.}
\label{OAM}
\end{center}
\end{figure}


Eq.\ref{SOI} shows the spin-to-orbital AM conversion in an inhomogeneous anisotropic medium (q-plate). Thus, the emerging wave is uniformly right and left circularly polarized (RCP/LCP) $\mathbf{E^{RCP/LCP}_{in} = [~1, ~~\mp i~~]^T}$ (as we would expect for an HWR), but in addition to that, it acquires an azimuthal phase factor $exp(\pm 2iq\phi)$. Here, $\pm 2q$ is the topological charge of the beam emerging from the q plate. In our experiment, as mentioned earlier, we used a q-plate with charge $q=1/2$ having a linear retardance of $\delta= \pi $. The q-plate generated the first-order $LG_{01}$ mode (or vortex beam) and converted the input helicity from LCP/RCP to RCP/LCP and vice versa, due to the spin-orbit interaction of light \cite{marrucci2006optical,fu2019spin}. Half of the intensity of the RCP/LCP-vortex beam emerging from the q-plate was transmitted through the beam splitter BS3, while the other half was reflected from BS3 (see Fig.~\ref{OAM Schematic}). The HWP within the RLI changed the helicity of the reflected beam from the RCP/LCP vortex ($\ell=1$) beam to the LCP/RCP vortex ($\ell=1$) beam. Subsequently, the first time the circulating LCP/RCP vortex beam passed through the q-plate, spin-to-orbit conversion occurred again, resulting in an increase/decrease of the vortex beam's OAM value by $\pm 1$ unit of $\hbar$ with RCP/LCP helicity. Consequently, the RCP/LCP vortex beam now carried an OAM of order $\pm 2$. This process repeats iteratively, with the OAM value increasing/decreasing by one order after each complete circulation of the beam within the RLI. Theoretically, we assume that the beams are circulated an infinite number of times in the RLI. Therefore, the value of OAM ($\ell$) ranges from $-\infty$ to $+\infty$, with the exception of zero. For the LCP and RCP input Gaussian beams, the values of OAM are positive and negative, respectively. 

We can compute the OAM series using conventional Jones matrix algebra for each cycle of the beam within the interferometer loop as follows: 
$$\left(\begin{array}{l}
E_x^i\\
E_y^i
\end{array}\right)=E_0\left(\begin{array}{c}
1 \\
\pm i
\end{array}\right)$$
\begin{equation}
\left(\begin{array}{l}
E_x^0 \\
E_y^0
\end{array}\right)=E_0\left(\frac{1}{4} e^{ \pm i \phi}+\left(\frac{1}{4}\right)^2 e^{ \pm 2 i \phi}+\left(\frac{1}{4}\right)^3 e^{ \pm 3 i \phi}+\cdots\right)\left(\begin{array}{c}
1 \\
\mp i
\end{array}\right)
\label{OAM series}
\end{equation}

\begin{center}
\begin{figure}
\includegraphics[width=\textwidth]{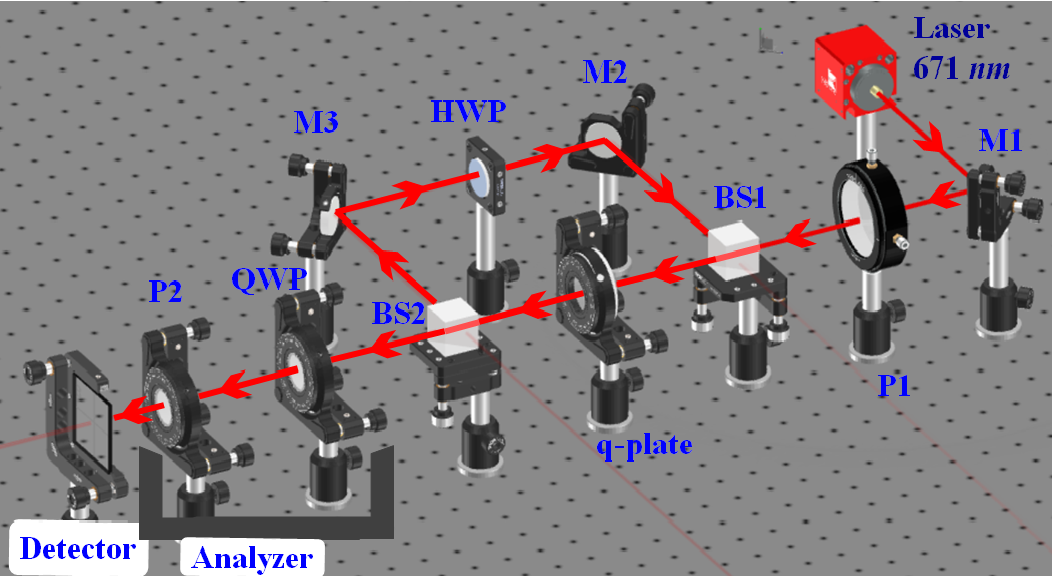}
\caption{Schematic diagram of our experimental setup used to produce a series of complex vector beams, where M1, M2, M3, M4 and M5 are the mirrors, P1 and P2 are wire grid polarizers, QWP1 is a quarter-wave plate, HWP is a half-wave plate, q-p is a q-plate, BS1 and BS2 are the 50-50 non-polarizing beam splitters.}
\label{multipole schematic}
\end{figure}
\end{center}

Here $\vec{E}^{i}=E_x^i \hat{e}_x+ E_y^i \hat{e}_y$, $\vec{E}^{0}=E_x^0\hat{e}_x + E_y^0\hat{e}_y$ denote the input and output electric fields, respectively, and E$_{0}$ is the amplitude of the input electric field. In Fig.~\ref{OAM}(a)-(f), we numerically computed the forked pattern, which is a combination of the phase distribution of $LG (\ell$) mode plus a linear phase ramp of Gaussian beam that creates a forked pattern, which shows the  OAM state $\ell=-3$ to $\ell=+3$ of the LG beam, respectively. By applying the Jones vector algebra on each optical component within the RLI  (see Fig.~\ref{OAM Schematic}) for an input LCP/RCP Gaussian beam, we obtained Eq.~\ref{OAM series} as a vortex beam series (or OAM series) at the output of RLI. With each complete circulation of the beams within the RLI, we had intensities ($I_{o} \sim {E_{0}^{2}}$) associated with OAM states, such as $(I_{o}/4)~ e^{\pm i \phi}$, $(I_{o}/16)~ e^{\pm 2 i \phi}$, and $(I_{o}/64)~ e^{\pm 3 i \phi}$ in the first, second, and third order of beam spots, respectively. As the beam circulated within the RLI, its intensity decreased by 1/4 of the previous intensity, while the topological charge (OAM) increased/decreased by one unit. For projecting out the values of the OAM states of different orders on our detector - we slightly misaligned the interferometer with the help of mirrors M4 or M5 (see Fig.~\ref{OAM Schematic}); as a result, we obtained different orders of beam spots, which were slightly shifted in the transverse plane after BS3. We performed an interference of the RCP/LCP vortex beam from path1 and the circularly polarized (RCP/LCP) Gaussian beam from path2 by using BS4 before the detector (see Fig.~\ref{OAM Schematic}). In addition, in the reference Gaussian beam path (path2), we employed a variable ND filter to achieve intensity matching among the different orders of vortex beams in path1. In Fig.~\ref{OAM} (g)-(l) and (m)-(r), we observe OAM values ranging from $\ell=-3$ to $\ell=+3$, which correspond to interference patterns of helical LG ($\ell \in [-3, 3]$) modes emerging from the q-plate ($q=1/2$) in the RLI. The corresponding interference patterns resulted from the superposition of the circularly polarized LG modes with a reference circularly polarized Gaussian beam. These patterns were observed under two different wavefront: plane (employing the collimated beam from the laser) and spherical (achieved by focusing the collimated beam by a lens). Figures on the left Fig.~\ref{OAM} [(g),(h),(i), (m), (n) and (o)], represent a right-circular input polarization (RCP), while those on the right, Fig.~\ref{OAM} [(j),(k),(l), (p), (q), and (r)], represent a left-circular input polarization (LCP). However, in Figs.~\ref{OAM} (g) (or (m)) and (l) (or (r)), we observed the -3 and +3 OAM orders, respectively. These orders appear slightly shifted from each other due to the challenge of aligning all three phase singularity points of the vortex beam. Subsequently, the intensity of the beams rapidly decreased, and the beams also stopped retracing the same path within the RLI due to the misalignment necessary to project them out separately. As a result, it is an impossible exercise to achieve a superimposition all the singularity points of the vortex beams experimentally.

\begin{center}
\begin{figure}[!h]
\includegraphics[width=\textwidth]{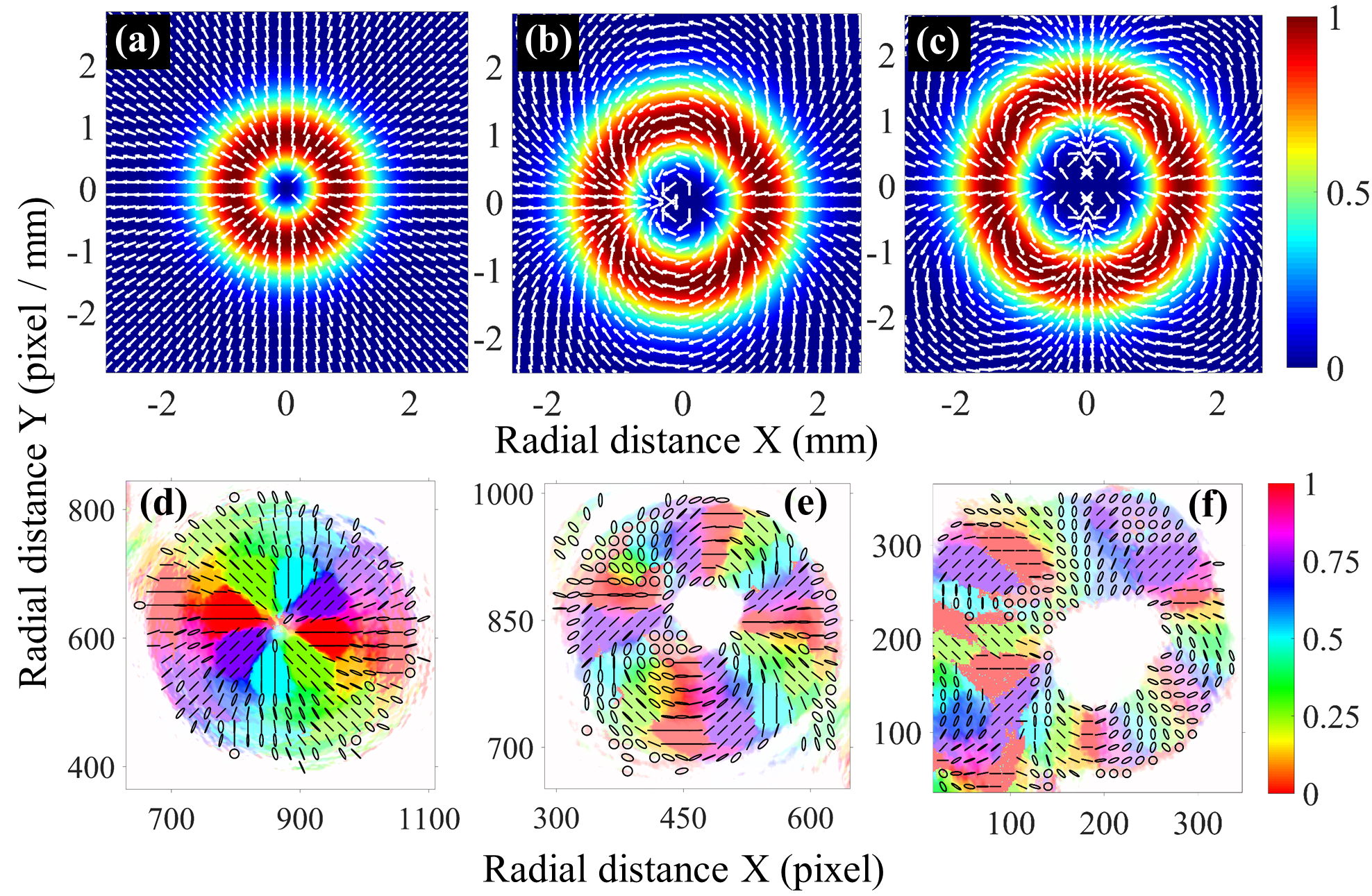}
\caption{ The upper panels labelled (a)-(c) depict the numerically calculated polarization states of the vector beam, specifically highlighting the radial, dipole, and quadrupole-like polarization configurations, respectively, each of which exhibits a polarization singularity at the center of the beam. Meanwhile, the lower panels designated (d), (e), and (f) illustrate the experimental measurements of the polarization states for the vector beam's radial, dipole, and quadrupole-like polarization configurations, respectively. These measurements were obtained by inputting a horizontally polarized Gaussian beam into an RLI (see Fig.~\ref{multipole schematic}) and analyzing the resulting polarization states using Stoke vector analysis.}
\label{multipole}
\end{figure}
\end{center}
\section{Generating a vector beam having a multipole-like polarization distribution using the RLI}

In this set of experiments (configuration shown in Fig.~\ref{multipole schematic}), we used linearly x-polarized light as input. The combination of the q-plate with a half-wave plate (HWP) in the RLI produced vector beams with a more complex polarization distribution. However, with each cycle of the beams within the interferometer loop, the size of the polarization (or intensity) singularity region gradually increased as the polarization state evolved. Ultimately, this process led to the generation of a vector beam that was the superposition of a series of vector beams. Once again, we assume that the beams are circulated an infinite number of times within the RLI. 

For an input beam that is x-polarized $E^{x-pol}_{input} = [~1, ~0~]^T$, we can write the expression of the output electric field as:
\begin{equation}
\begin{aligned}
& {\left[\begin{array}{l}
E_x^0 \\
E_y^0
\end{array}\right]=E_0\left(\frac{1}{2^2}\left[\begin{array}{c}
\cos \phi \\
\sin \phi
\end{array}\right]+\frac{1}{2^4}\left[\begin{array}{c}
\cos 2 \phi \\
\sin 2 \phi
\end{array}\right]+\frac{1}{2^6}\left[\begin{array}{c}
\cos 3 \phi \\
\sin 3 \phi
\end{array}\right]+\cdots\right)} \\
\label{Polarization}
\end{aligned}
\end{equation}
where, E$_{0}$ is the amplitude of the input beam and $\phi$ is the azimuthal angle with respect to the lab frame. Theoretically, using the Jones matrix algebra, we calculated the series of vector beams. In Eq.~\ref{Polarization}, the first term of the output electric field represents radial (monopole) polarization. The second term corresponds to a dipole-like polarization, the third to a quadrupole-like polarization, and so on. Hence, the output electric field (Eq.~\ref{Polarization}) consisted of the superposition of a series of vector beams that have a multipole-like polarization distribution, with intensity decreasing by a factor of 1/4 for each increasing order.

Generally, a q-plate having a fixed q=1/2 value can generate only a first-order vector beam, while higher-order vector beams can be generated by the addition of two or more q-plates. However, it is well known that inserting an HWP between the two q-plates of the same or different charges results in the addition of the charges of the q-plates. Following this, we inserted an HWP inside the RLI so that the effective charges of the single q-plate increased by the number of times the beams circulate within the RLI, so as to generate higher-order vector beams. In Fig.~\ref{multipole schematic}, we again deliberately misaligned the interferometer with the help of either mirror M2 or M3 so that we could project out the different order of vector beams at the detector. Using Stokes vector analysis, we then measured the polarization states of these vector beams. To characterize the polarization state of each order (up to three orders) of the vector beams, we collected six sets of data using the analyzer system depicted in Fig.~\ref{multipole schematic}. Using a polarimetry code, we plotted the polarization ellipse for the first three orders of the vector beams described in Eq.~\ref{Polarization}. In Figs.~\ref{multipole} (a), (b), and (c), we numerically computed the intensity and polarization distributions of the first (monopole), second (dipole), and third terms (quadrupole) of the output electric field of Eq.~\ref{Polarization}, respectively. In Figs.~\ref{multipole} (d), (e), and (f), experimentally, we then measured the polarization distribution of the first, second, and third order of vector beams, respectively. Our measurements of the polarization distributions of the first and second terms of the output electric field as mentioned in Eq.~\ref{Polarization} were very accurate, while the polarization distributions of the third term do not match very well with the numerically computed distributions due to the limitations in the accurate alignment of all three polarization singularity points. The beams containing the higher-order multipoles are no longer confined in the RLI, so the superposition (by accurate alignment) of their polarization singularity points is impossible.

\section{Conclusion}
In conclusion, we have developed a novel RLI that confines light in a rectangular path geometry and thereby leads to a host of interesting and intriguing applications by exploiting the scalar and vector superposition of light. In the scalar superposition applications, we computed the sum of numerous converging mathematical geometric series, which converge to one or less than one, by varying the intensity circulating in the interferometer by using simple reflective (or transmissive) optics (BS) or more complex polarization control optical devices (a combination of HWP and PBS). The convergence values were determined at the speed of light, with the response time of detection bandwidth finally determining the computation speed. Our accuracy was chiefly limited by alignment issues into the interferometer and varied between 90\%-98\% once those losses were taken into account. 

In the cases of vector superposition, first, we physically generated a vector-vortex (OAM) beam series by introducing a q-plate and an HWP into the RLI. As the beam circulated within the RLI, the effective charge of the q-plate increased, resulting in large SOI effects for an input circularly polarized Gaussian beam. However, while the topological charge (OAM) of the beam increased/decreased by one unit with each pass in the interferometer (leading to an increase in the extent of the phase singularity at the beam central region),  the corresponding intensity decreased by 1/4 for each subsequent charge. In theory, we generated a beam containing an infinite orthogonal basis set of OAM for light, excluding zero. However, experimentally, we were able to project out OAM values from $\ell=-3$ to $\ell=+3$ by misaligning the optics slightly. Higher-order beams were no longer confined within the interferometer due to this misalignment. 

Furthermore, with an input of linearly polarized light, our configuration (RLI) enabled the physical generation of a vector beam with a complex poliarization state that was the linear superposition of polarization multipoles. The size of the polarization singularity region gradually increased as the polarization distribution evolved inside the interferometer. However, experimentally, using Stoke vector analysis, we were able to measure the polarization distribution of the first (monopole), second (dipole), and third-order (quadrupole) vector beams; after that, the beams diverged from the interferometer. Our configuration also paves the way for observing the helicity-dependent orbital Hall effect and large enhancement of the geometrical Pancharatnam–Berry phase, which we plan to report in future work. In addition, we also envisage extending the RLI to a polygonal loop interferometer (PLI) that can confine light within an N-order polygonal geometry for determining the sum of any variable geometric series by employing different beam splitters having variable reflection/transmission ratios at the vertices of the polygon.

\section{Acknowledgement}

The authors acknowledge the SERB, Department of Science and Technology, Government of India (Project No. EMR/2017/001456) and IISER Kolkata IPh.D fellowship for research. We also acknowledge Sramana Das for her help in experiments.



\begin{appendix}

\section{Sum of mathematical series using the RLI}

\begin{figure}
    \begin{center}
\includegraphics[width=0.8\textwidth]{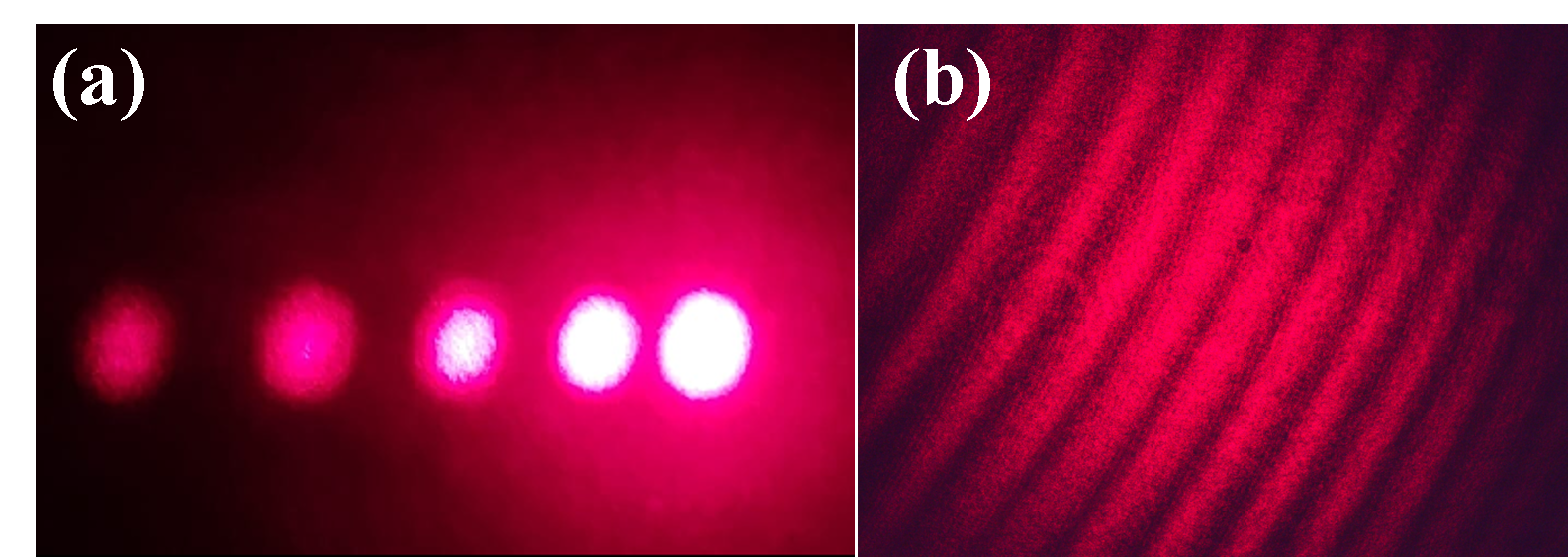}
\caption{(a) Multiple beam spots are obtained on the screen (at detector D2, see Fig.~\ref{series sch} (a)) just by doing misalignment of the interferometer with the help of either mirrors M1 or M2 of RLI. (b) An interference pattern is obtained on the screen when all beam spots are matched at one point.}
\label{RLI}
\end{center}
\end{figure}

By deliberately misaligning the interferometer using the mirrors M1 or M2 of the RLI, we obtained multiple beam spots on the screen at detector D2 (or D1) (see Fig.~\ref{series sch}). The brightest beam spot in Fig.~\ref{RLI} (a) was the first-order output from the BS2, whose intensity corresponded to the first term of Eq.~\ref{I2}. The second, third, fourth, and fifth beam spots corresponded to the output of the first, second, third, and fourth passes of the beam in the loop of RLI, whose intensity corresponded to the second, third, fourth, and fifth terms of Eq.~\ref{I2}, respectively. As the beam circulated in the loop, its intensity decreased by a factor of around 1/4 times with each pass. Note that we could not obtain higher (more than five) order beam spots because of the divergence of the beam from the RLI. The interference pattern, as shown in Fig.~\ref{RLI} (b), emerged when all the beam spots coincide at one point, ensuring optimal alignment of the interferometer. That is why we call this setup Fig.~\ref{series sch} a rectangle loop interferometer (RLI).  

\subsection{{Sum of a converging geometric series ($ \lesssim 1$) using the evolution of light intensity in the RLI}}

We measured the input and output intensity as well as before and after at all optical components used in the setup depicted in Fig.\ref{series sch} (a) and (b) so that we could calculate the series sum I$_{1}$ and I$_{2}$ ( see Eqs. \ref{I1} and ~\ref{I2}) and the optical losses.

\begin{equation}
\begin{array}{|c|c|c|c|c|c|c|c|c|c|}
\hline \text { Points } & \mathbf{I}_{\mathbf{a}} & \mathbf{I}_{\mathbf{b}} & \mathbf{I}_{\mathbf{c}} & \mathbf{I}_{\mathbf{c c}} & \mathbf{I}_{\mathbf{d}} & \mathbf{I}_{\mathbf{d d}} & \mathbf{I}_{\mathbf{e}} & \mathbf{I}_{\mathbf{f}} & \mathbf{I}_{\mathbf{g}} \\
\hline \text { Power }(\mathbf{m W}) & 2.980 & 1.402 & 1.302 & 1.703 & 0.600 & 0.750 & 0.665 & 0.630 & 0.599 \\
\hline
\end{array}
\label{table1}
\end{equation}

 where, $\mathbf{I}_{\mathbf{a}}$, $\mathbf{I}_{\mathbf{b}}$, $\mathbf{I}_{\mathbf{e}}$, $\mathbf{I}_{\mathbf{f}}$ are the intensity at the points mentioned in Fig.~\ref{series sch} (a). $\mathbf{I}_{\mathbf{c}}$ and $\mathbf{I}_{\mathbf{d}}$ are the intensity at points c and d after blocking one arm of the RLI, respectively. However, $\mathbf{I}_{\mathbf{cc}}$ and $\mathbf{I}_{\mathbf{dd}}$ are the total intensity measured at detector D1 and D2, respectively.

Using table \ref{table1}, we calculated the reflection and transmission coefficient of beam splitters.
\begin{equation}
r_1=\frac{I c}{I a} ; r_2=\frac{I_d}{I_b} ; t_1=\frac{I_b}{I_a} ; t_2=\frac{I_d}{I_b}
\label{rt}
\end{equation}

where, $r_1$, $r_2$, $t_1$ and $t_2$ are the reflection and transmission coefficient of BS1 and BS2, respectively. Using \ref{table1} and \ref{rt}, the coefficients are given as

$r1=0.4369; r2= 0.4705; 
t1=0.4705; t2= 0.4280
$
The combined reflection coefficient of mirrors is $r_{m} =I_{g}/I_{e}=0.900$.
Using the above reflections and transmission coefficients, we calculate the sum of the series I$_{1}$, I$_{2}$ ( see Eqs.\ref{I1} and \ref{I2}) and \% error at the detector D1 and D2, respectively as


\begin{equation}
\begin{tabular}{|c|c|c|c|c|}
\hline $\mathbf{I}_{\text {theory }}$ & $\mathbf{I}_{\mathbf{e x p}}$ & $\begin{array}{c}\text { Error } \\
\text { (Measurable) }\end{array}$ & $\begin{array}{c}\text { Error \% } \\
\text { (Measurable) }\end{array}$ & $\begin{array}{c}\text { Error \% } \\
\text { (Not measurable) }\end{array}$ \\
\hline 1.986 & 1.703 & 0.251 & 12.63 & 1.62 \\
\hline
\end{tabular}\\
\label{table2}
\end{equation}

\begin{equation}
\begin{tabular}{|l|c|c|c|c|}
\hline \multicolumn{1}{|c|}{$\mathbf{I}_{\text {theory }}$} & $\mathbf{I}_{\mathbf{e x p}}$ & $\begin{array}{c}\text { Error } \\
\text { (Measurable) }\end{array}$ & $\begin{array}{c}\text { Error \% } \\
\text { (Measurable) }\end{array}$ & $\begin{array}{c}\text { Error \% } \\
\text { (Not measurable) }\end{array}$ \\
\hline 0.993 & 0.750 & 0.180 & 18.126 & 6.345 \\
\hline
\end{tabular}
\label{table3}
\end{equation}

where, Tables~\ref{table2} and \ref{table3} correspond to series I$_{1}$ and I$_{2}$ (see Eqs.\ref{I1} and \ref{I2}) at detectors D1 and D2, respectively (see Fig. \ref{series sch} (a)). Intensity losses through optical components enhanced the error by around 12\% at detector D1 and 18\% at detector D2, respectively. However, even after we took these errors into account in the measured sum of the series, there remained additional losses -- approximately 1.6\% at D1 and 6.3\% at D2, respectively. These could can be attributed to the paraxial beam approximation and alignment issues. These losses cannot be quantified by direct measurement. \\

\begin{center}
\begin{figure}
\includegraphics[width=\textwidth]{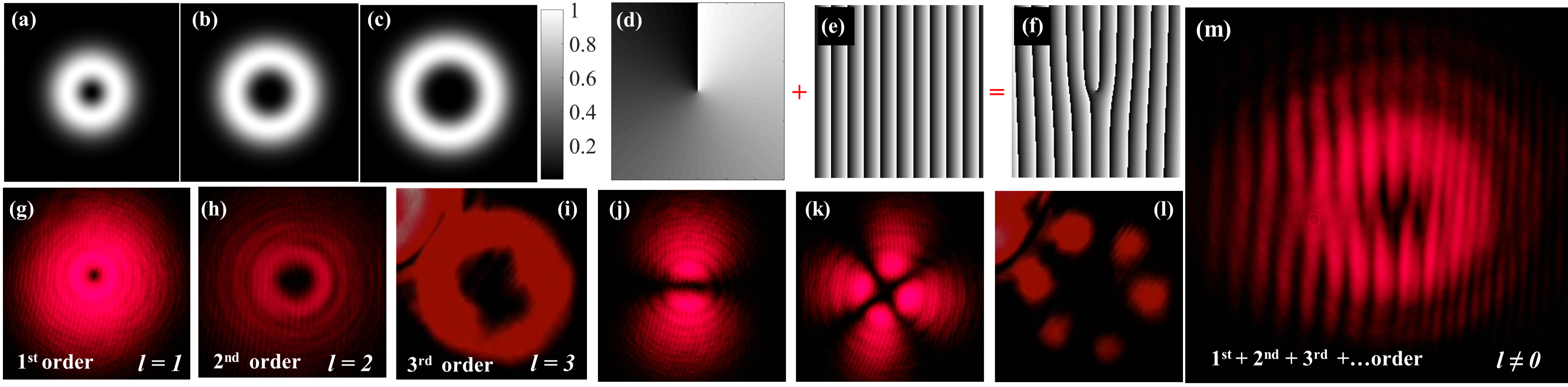}
\caption{ (a), (b), and (c) represent the numerically computed intensity profiles for the LG$_{01}$, LG$_{02}$, and LG$_{03}$ modes, respectively. (d) Demonstrate the phase distribution of the LG$_{01}$ mode, (e) displays a characteristic linear phase ramp of the Gaussian mode, and (f) showcases the distinctive fork pattern produced through a linear combination of the phase distribution of the LG$_{01}$ mode and the linear phase ramp of the Gaussian mode. On the other hand, (g), (h), and (i) show the experimentally obtained images of the LG$_{01}$, LG$_{02}$ and LG$_{03}$ modes, respectively. (j), (k) and (l) illustrate the experimentally obtained images of the LG$_{01}$, LG$_{02}$ and LG$_{03}$ modes, respectively, after passing through a polarizer. (m) Illustrates the self-interference of a different order of LG modes without deviating the beam from the RLI, resulting in a non-zero value of OAM ($l\not = 0$).}
\label{LG beam exp}
\end{figure}
\end{center}

\subsection{Sum of a converging geometric series converging to $\sim 1$ using the evolution of polarization-dependent light intensity in the RLI}
Tables~\ref{table4} and \ref{table5} correspond to the series Eq.~\ref{Ip} (I$_{p}$). In Fig.~\ref{series sch} (b), the intensity at points a, b, c, and d are displayed in  Tables~\ref{table4} and \ref{table5}.

\begin{equation}
\begin{array}{|c|c|c|c|c|c|}
\hline \text { Points } & \mathbf{I}_{\mathbf{a}} & \mathbf{I}_{\mathbf{b}} & \mathbf{I}_{\mathbf{c}} & \mathbf{I}_{\text {block }} & \mathbf{I}_{\text {abs }} \\
\hline \text { Power }(\mathbf{m W}) & 23.640 & 22.290 & 19.420 & 0.424 & 4.1313 \\
\hline
\end{array}
\label{table4}
\end{equation}

where, $\mathbf{I}_{\mathbf{a}}$, $\mathbf{I}_{\mathbf{b}}$, and $\mathbf{I}_{\mathbf{c}}$ are the intensity at points a, b, and c mentioned in Fig.~\ref{series sch} (b), respectively. $\mathbf{I}_{\mathbf{block}}$ is the intensity after blocking one arm of RLI so that the beam can no longer rotate in the loop of RLI in Fig.~\ref{series sch} (b) (i.e. intial reflected intensity from the PBS). At the end of the calculation, we subtract $\mathbf{I}_{\mathbf{block}}$ from I$_{p}$ ($\mathbf{I}_{\mathbf{block}}$ is an error). $\mathbf{I}_{\mathbf{abs}}$ is the total absorbed intensity by PBS, HWP and mirrors of Fig.~\ref{series sch} (b). \\

\begin{tabular}{|c|c|c|c|c|c|c|c|}
\hline HWP $(\theta)$ & $0^{\circ}$ & $10^{\circ}$ & $15^{\circ}$ & $20^{\circ}$ & $30^{\circ}$ & $40^{\circ}$ & $45^{\circ}$ \\
\hline \begin{tabular}{c} 
Points $\left(I_d\right)$ \\
Power (mW)
\end{tabular} & 19.120 & 17.390 & 16.880 & 16.740 & 16.510 & 16.460 & 16.400 \\
\hline $\mathrm{I}_{\mathrm{p}}^{\exp }$ & 0.9835 & 0.910 & 0.888 & 0.883 & 0.873 & 0.871 & 0.868 \\
\hline
\end{tabular}

\begin{equation}
\begin{array}{|c|c|c|c|c|c|c|}
\hline \text { HWP }(\theta) & 50^{\circ} & 60^{\circ} & 70^{\circ} & 75^{\circ} & 80^{\circ} & 90^{\circ} \\
\hline \begin{array}{c}
\text { Points }\left(I_d\right) \\
\text { Power }(\mathrm{mW})
\end{array} & 16.510 & 16.680 & 17.350 & 17.930 & 19.090 & 19.060 \\
\hline \mathbf{I}_{\mathrm{p}}^{\exp } & 0.873 & 0.880 & 0.908 & 0.933 & 0.982 & 0.981 \\
\hline
\end{array}
\label{table5}
\end{equation}

where, ${I}_{{d}}$ is the intensity measured at the point d mentioned in Fig.~\ref{series sch} (b) and $ \mathbf{I}^{\mathrm{exp}}_{\mathrm{p}}$ is the value of series I$_{p}$ (see Eq.~\ref{Ip}) for different values of $\theta$.

\section{Higher-order vortex modes (OAM) generation using a single q-plate}

In Figs.~\ref{LG beam exp} (a)-(c), show the numerically computed intensity profiles for the LG$_{01}$, LG$_{02}$, and LG$_{03}$ modes, respectively (see Eq.~\ref{LG beam theo}). Fig.~\ref{LG beam exp} (d) illustrates the phase distribution of the LG$_{01}$ mode (see Eq.~\ref{LG beam theo}). Fig.~\ref{LG beam exp} (e) shows a linear phase ramp characteristic of the reference Gaussian mode of path 2 (see Fig.~\ref{OAM Schematic}). The numerically computed fork pattern, as shown in Fig.~\ref{LG beam exp} (f), is the linear combination of the phase distribution of the LG$_{01}$ mode and the linear phase ramp of the Gaussian mode. Figs.~\ref{LG beam exp} (g), (h) and (i) are the experimentally obtained images of the LG$_{01}$, LG$_{02}$, and LG$_{03}$ modes, respectively, just by slightly misaligning the RLI (see Fig.~\ref{OAM Schematic}).  At the same time, Figs.~\ref{LG beam exp} (j), (k) and (l) exhibit 2, 4, and 6 lobes, respectively, representing the LG$_{01}$, LG$_{02}$, and LG$_{03}$ modes, after passing through the polarizer. As the beam circulated in the loop of RLI, we obtained different orders of beam spots with decreasing order of intensity. The first, second and third-order beam spots contained circularly polarized vortex beams having topological charges of one, two and three, respectively, at the output of RLI. Fig.~\ref{LG beam exp} (r) shows the self-interference of all possible orders of the LG mode with good alignment inside the RLI (i.e. no misalignment); as a result, we get a non-zero value of the OAM ($l\not = 0$), which cannot be split into individual topological states.


\end{appendix}




\nolinenumbers

\end{document}